# On the spin parameter of dark-matter haloes

**Matthias Steinmetz and Matthias Bartelmann**

Max-Planck-Institut für Astrophysik, Postfach 1523, D–85740 Garching, FRG

8 March 1994

**Abstract.** The study by White (1984) on the growth of angular momentum in dark haloes is extended towards a more detailed investigation of the spin parameter $\lambda \equiv L\sqrt{E}/GM^{2.5}$. Starting from the Zel'dovich approximation to structure formation, a dark halo is approximated by a homogeneous ellipsoid with the inertial tensor of the (highly irregular) Lagrangian region $\Upsilon$ from which the dark halo forms. Within this approximation, an expression for the spin parameter can be derived, which depends on the geometry of $\Upsilon$, the cosmological density parameter $\Omega_0$, the overdensity of the dark halo, and the tidal torque exerted on it. For Gaussian random fields, this expression can be evaluated statistically. As a result, we derive a probability distribution of the spin parameter which gives $\lambda \simeq 0.07^{+0.04}_{-0.05}$, consistent with numerical investigations. This probability distribution steeply rises with increasing spin parameter, reaching its maximum at $\lambda \simeq 0.025$. The 10 (50,90) percentile values are $\lambda = 0.02$ (0.05,0.11, respectively). There is a weak anticorrelation of the spin parameter with the peak height $\nu$ of the density fluctuation field $\lambda \propto \nu^{-0.29}$. The dependence on $\Omega_0$ and the variance $\sigma$ of the density-contrast field is very weak; there is only a marginal tendency for the spin parameter to be slightly larger for late-forming objects in an open universe. Due to the weak dependence on $\sigma$, our results should be quite generally applicable and independent on the special kind of the fluctuation spectrum.

**Thesaurus codes:** 11.06.1, 11.06.2, 11.11.1, 12.03.4, 12.04.1

## 1 Introduction

We adopt the widely accepted view that galaxies acquire angular momentum from the tidal torques of surrounding matter (Hoyle 1949, Peebles 1969, Doroshkevich 1970, White 1984). Most of the angular momentum $L$ is gained during the linear phase of the collapse, while an overdense region is expanding slower than the Hubble flow (i.e., collapsing in comoving coordinates) and angular momentum is growing linearly with time ($L \propto t$, White 1984). After the turnaround point is reached and the collapse becomes nonlinear, further gain of angular momentum is strongly suppressed because it becomes increasingly difficult for the tidal field to act on the shrinking radius of the forming dark halo. Numerical simulations by White (1984) support this picture for the growth of the galactic angular momentum at least in a statistical sense. Numerical simulations by Efstathiou & Jones (1979), Barnes & Efstathiou (1987) and Warren et al. (1992) showed an average spin parameter $\lambda \equiv L\sqrt{E}/GM^{2.5}$ of about $0.05\ldots 0.07$, where $L$, $E$, and $M$ are the total angular momentum, energy, and mass of the dark halo, respectively, and $G$ is the gravitational constant. Only 10% of the dark matter haloes in such simulations have a spin parameter of either less than 0.02 or more than 0.11. Such low values for the spin parameter indicate that the dark haloes are supported by the velocity dispersion of their constituents, and that the flattening of the haloes is a consequence of an anisotropic velocity dispersion and not of rotation. Furthermore, the spin parameter seems to be nearly independent on the peak height and the spectral index of the fluctuation spectrum.

Up to now, several analytical studies investigated the growth of angular momentum in hierarchical scenarios, most of them within the CDM cosmogony in $\Omega_0 = 1$ model universes (Hoffman 1986, 1988, Ryden 1988, Heavens & Peacock 1988, Quinn & Binney 1992). These investigations yield spin parameters on the order of 0.07, consistent with $N$-body simulations. The more detailed studies of Ryden (1988) and Heavens & Peacock (1988) employ the peak formalism of Bardeen et al. (1986). Furthermore, Hoffman (1986) argued that the average spin parameter should be anticorrelated with the peak height of density fluctuations. Based on dimensional arguments, White (1994) suggested that the spin parameter in an open universe ($\Omega_0 < 1$) should be systematically lowered by $\Omega^{0.6}(t)$, where $\Omega(t)$ has to be taken at turnaround time. However, a more systematic study of the spin parameter and its distribution remains to be done.

In this paper, we investigate the statistical properties of the spin parameter in terms of global properties of Gaussian random fields; therefore, our results are independent of the shape of the power spectrum. In Sect.2, we derive expressions for the mass, the angular momentum, the energy, and, consequently, for the spin parameter of a dark halo. In Sect.3, the expression for the spin parameter is statistically analyzed, assuming that the primordial density fluctuations can be described by a Gaussian random field. The resulting probability distribution is then discussed in terms of the shape of the dark halo, the peak height, and the background cosmology. The results are then compared to the outcome of $N$-body simulations.

## 2 Fundamental parameters of dark haloes

In this section, we derive expressions for mass, angular momentum, energy, and spin parameter of dark haloes within the framework of linear theory. The dimensionless spin parameter is defined by

$$\lambda = \frac{L\sqrt{|E|}}{GM^{5/2}} \,. \tag{1}$$

Note that its dependence on the total energy of the system is rather weak.

### 2.1 Linear theory and Zel'dovich approximation

For sufficiently early times, the evolution of density perturbations $\delta(\boldsymbol{r}, t) = [\varrho(\boldsymbol{r}, t) - \varrho_\mathrm{b}]/\varrho_\mathrm{b}$ is well described by linear theory. $\varrho(\boldsymbol{r}, t)$ is the density at a given point $(\boldsymbol{r}, t)$ in spacetime, and $\varrho_\mathrm{b}$ is the mean background density. The density contrast $\delta$ fulfils Poisson's equation, $\Delta \Phi = -\delta$, $\Phi$ being the peculiar gravitational potential. In Eulerian formulation, the comoving position of a perturbation in space is fixed and only its amplitude changes,

$$\delta(t) = b(t)\delta(t_\mathrm{i}) \,, \tag{2}$$

where $b(t_{\rm i})$ is normalized to unity. In an Einstein-de Sitter universe, $b(t) = a(t)$, $a(t)$ being the dimensionless cosmological expansion function, normalized to unity at $t = t_{\rm i}$. For an open universe ($\Omega_0 < 1$), the linear growth factor $b(t)$ can be approximated to within a few percent by

$$\delta(z) \simeq \delta(z_{\rm i}) \frac{1 + \tilde{\Omega} z_{\rm i}}{1 + \tilde{\Omega} z} , \qquad (3)$$

with

$$\tilde{\Omega} \equiv \frac{2.5 \Omega_0}{1 + 1.5 \Omega_0} \qquad (4)$$

(Shandarin et al. 1983), $z = (1 + z_{\rm i})/a(t) - 1$ being the redshift corresponding to time $t$.

The accuracy of linear theory can be improved if particle trajectories are considered instead of the density contrast. This technique is known as Zel'dovich approximation (Zel'dovich 1970) and is most easily understood in the Lagrangian formulation of perturbation theory (Buchert 1992). Let $r_j$, $j \in \{1, 2, 3\}$, be physical particle coordinates starting at Lagrangian coordinates $q_j$. Then,

$$r_j(q_j, t) = a(t) [q_j + (b(t) - 1) \Phi_{,j}(\boldsymbol{q})] , \qquad (5)$$

where the comma followed by an index $j$ indicates the partial derivative with respect to $q_j$.

## 2.2 Angular momentum

Consider now a fiducial trajectory $r_j(q_j, t)$ which, for instance, could be the trajectory of the center-of-mass of the mass distribution to be considered. For a neighbouring trajectory, we then have

$$\delta r_j = \frac{\partial r_j}{\partial q_k} \delta q_k = a(t) [\delta q_j + (b(t) - 1) \Phi_{,jk} \delta q_k] , \qquad (6)$$

where the sum convention is implied. Since the particle velocity is

$$v_j = \dot{r}_j = \dot{a}(t) q_j + \frac{d}{dt}[a(t)(b(t) - 1)] \Phi_{,j} , \qquad (7)$$

the velocity of the particle considered relative to the fiducial particle is

$$\delta v_j = \dot{a}(t) \delta q_j + \frac{d}{dt}[a(t)(b(t) - 1)] \Phi_{,jk} \delta q_k . \qquad (8)$$

Hence, the angular momentum of a volume element $dV$ with density $\varrho$ at distance $\delta r_j$ relative to the fiducial particle is

$$dL_j = \varrho dV \epsilon_{jkl} \left[ a(t) \frac{d}{dt}[a(t)(b(t) - 1)] \Phi_{,ln} \delta q_k \delta q_n + a(t)(b(t) - 1) \dot{a}(t) \Phi_{,km} \delta q_l \delta q_m \right] , \qquad (9)$$

where $\epsilon_{jkl}$ is the totally antisymmetric tensor with $\epsilon_{123} = 1$. Integrating over the volume of interest $\Upsilon$, we obtain after changing the summation indices and accounting for the antisymmetry of $\epsilon_{jkl}$

$$L_j = a^2(t) \dot{b}(t) \epsilon_{jkl} I_{kn} \Phi_{,ln} , \qquad (10)$$

where we have defined the inertial tensor of $\Upsilon$

$$I_{jk} \equiv \int_\Upsilon dV \; \varrho(t_\mathrm{i}) \delta q_j \delta q_k \; . \tag{11}$$

This result was first derived by White (1984), who also discusses Eq.(10) in detail. Note that the origin of the angular momentum is the shear flow of the cosmic medium. In the Zel'dovich approximation, motion is restricted to irrotational flows. This guarantees the existence of the potential $\Phi$ whose gradient is the velocity field and whose source can, in linear approximation, be identified with the density contrast field $\delta$. Note also that we use the Zel'dovich approximation only for the derivation of Eq.(10); it enters the derivation of the spin parameter only in this respect.

Now, we approximate the shape of $\Upsilon$ at initial time $t_\mathrm{i}$ by a triaxial homoeoid of density $\varrho$ with the same inertial tensor as $\Upsilon$. Let the principal axes $u_j$ be ordered such that $u_1 > u_2 > u_3$. We introduce the ellipticity parameter $e \equiv (u_2/u_1)$ and the prolateness parameter $p \equiv (u_3/u_1)$; then, $1 > e > p$. Rigorously, the $u_j$ depend on time; however, since we need the inertial tensor only at $t_\mathrm{i}$, we omit the time dependence and use the abbreviation $u_j \equiv u_j(t_\mathrm{i})$. After diagonalization, the inertial tensor reads

$$I_{jk} = \mathrm{diag}(\mu_j) \; , \; j \in \{1,2,3\} \; , \tag{12}$$

where

$$\mu_1 = \frac{M}{5} u_1^2 \; , \; \mu_2 = \mu_1 e^2 \; , \; \mu_3 = \mu_1 p^2 \; . \tag{13}$$

The total mass $M$ enclosed by the ellipsoid is

$$M = \frac{4\pi}{3} u_1^3 \varrho e p \; . \tag{14}$$

Referring the density to the background density $\varrho_\mathrm{b}$ of the underlying Friedmann-Lemaître model, we have at an initial time $t_\mathrm{i}$

$$\varrho(t_\mathrm{i}) = \varrho_\mathrm{b}(t_\mathrm{i})(1 + \delta_\mathrm{i}) \; , \tag{15}$$

where $\delta_\mathrm{i}$ is the initial density contrast of the volume. In terms of the initial cosmological density parameter $\Omega_\mathrm{i}$,

$$\varrho_\mathrm{b}(t_\mathrm{i}) = \frac{3 H_\mathrm{i}^2 \Omega_\mathrm{i}}{8\pi G} \; , \tag{16}$$

where $H_\mathrm{i}$ is the Hubble parameter of the background model at $t_\mathrm{i}$. Combining Eqs.(14), (15) and (16), we obtain

$$M = \frac{H_\mathrm{i}^2 \Omega_\mathrm{i}}{2G}(1 + \delta_\mathrm{i}) u_1^3 e p \; . \tag{17}$$

From Eq.(13), we get

$$\mu_1^3 = \frac{M^3}{125} u_1^6 \tag{18}$$

and, together with Eq.(17), this yields

$$M^{5/2} = \sqrt{\frac{125}{4}} \frac{H_\mathrm{i}^2 \Omega_\mathrm{i}(1+\delta_\mathrm{i})}{G} \mu_1^{3/2} e p \; . \tag{19}$$

## 2.3 Potential energy

A homoeoid with principal axes $u_j$ and density $\varrho$ has the potential energy

$$W = -\frac{8\pi^2}{15}G\varrho^2 u_1^5 \left[2e^2 p^2 \frac{\mathrm{F}(\theta,k)}{\sin\theta}\right],\tag{20}$$

see, e.g., Binney & Tremaine (1987, table 2.1). In Eq.(20), $\mathrm{F}(\theta,k)$ is the incomplete elliptic integral of first kind, with

$$\theta \equiv \arccos(p),\ k \equiv \sqrt{\frac{1-e^2}{1-p^2}}\ .\tag{21}$$

Together with Eq.(14), we obtain from Eq.(20)

$$W = -\frac{3}{5u_1}GM^2 \frac{\mathrm{F}(\theta,k)}{\sin\theta}\ .\tag{22}$$

The potential energy equals the total energy when the ellipsoid achieves its maximum extension. This neglects the rotational energy of the ellipsoid, which is, however, small compared to the potential energy, as will be confirmed below. A homogeneous sphere with density contrast $\delta_i$ expands until it reaches the radius

$$u_1^{\max} = u_1 \frac{1+\delta_i-\epsilon_i}{\frac{5}{3}\delta_i-\epsilon_i}\ ,\tag{23}$$

see, e.g., Bartelmann et al. 1993, Eq.(23); there, it was also shown that the collapse history of a homoeoid is, at least statistically, well approximated by the collapse timescale of a sphere. $\epsilon_i \equiv 1 - \Omega_i$ is the deviation of the initial density parameter from unity. That means that density fluctuations with $\delta_i \leq (3/5)\epsilon_i$ never collapse. Therefore, it is convenient to introduce the quantity

$$\gamma_i \equiv \frac{5}{3}\delta_i - \epsilon_i\ ,\tag{24}$$

i.e., density fluctuations with $\gamma_i > 0$ collapse, those with $\gamma_i \leq 0$ do not; since $\delta_i \ll 1$ and $\epsilon_i \ll 1$, $\gamma_i$ is also much smaller than unity. Approximating the maximum value of the largest principal axis of the ellipsoid by (23), we obtain for the total energy

$$E = -\frac{3G\gamma_i M^2}{5u_1}\frac{\mathrm{F}(\theta,k)}{\sin\theta}\ .\tag{25}$$

Moreover, from Eq.(13),

$$M^{1/2} = \frac{\sqrt{5\mu_1}}{u_1}\ ,\tag{26}$$

and therefore

$$E = -\frac{3G}{5}\frac{\gamma_i M^{5/2}}{\sqrt{5\mu_1}}\frac{\mathrm{F}(\theta,k)}{\sin\theta}\ ,\tag{27}$$

an expression for the energy which turns out convenient later on. Note that the approximation for the energy is the probable crudest in the investigation presented here. The resulting error, however, is relatively small because only $\sqrt{E}$ enters the definition of the spin parameter (see below).

### 2.4 The spin parameter

From Eqs.(10) and (12), we obtain for the angular momentum $L$ in the principal-axis system of the inertial tensor

$$L^2 = \left[a^2(t)\dot{b}(t)\right]^2 \sum_{j>k} \left[(\mu_j - \mu_k)^2 \Phi_{,jk}^2\right] ; . \tag{28}$$

Therefore, the square of the spin parameter (1) reads

$$\lambda^2 = \frac{6}{125} \left[a^2(t)\dot{b}(t)\right]^2 \frac{\gamma_i}{H_i^2} \frac{F(\theta,k)}{\sin\theta} \frac{\sum_{j>k}\left[(\mu_j-\mu_k)^2\Phi_{,jk}^2\right]}{\mu_1\sqrt{\mu_2\mu_3}} . \tag{29}$$

In somewhat more detail, the term $F(\theta,k)/\sin\theta$ reflects the change in gravitational binding energy compared to a homogeneous sphere, i.e., it is an intrinsic property of $\Upsilon$. The rightmost part of Eq.(29) describes the tidal interaction of $\Upsilon$ with the surrounding matter. Note that in the numerator the same power of $\mu_j$ appears as in the denominator; i.e., the term depends only on the axis ratios of $\Upsilon$. The term $\Phi_{,jk}$ is dimensionless and of the same order of magnitude as $\delta_j$. The term $(a^2\dot{b})^2/H_i^2$ measures the time evolution of the angular momentum in units of the initial Hubble parameter $H_i$. This term will be considered in the following subsection.

### 2.5 Dependence on time

The time-dependent factor in square brackets in Eq.(29) remains to be evaluated. To do so, we start from the definition of the 'velocity factor' $f$,

$$f \equiv \frac{a}{\dot{a}}\frac{\dot{b}}{b} \simeq \Omega^{0.6}(t) ; \tag{30}$$

see, e.g., Peebles [1993, Eqs.(5.116) and (5.120)]. With Eq.(30), the time-dependent factor in Eq.(29) can be written

$$a^2\dot{b} = a^2 H b f . \tag{31}$$

For sufficiently large collapse redshifts, we can employ the Einstein-de Sitter limit,

$$a(t) = \left(\frac{t}{t_i}\right)^{2/3} \quad , \quad b(t) \simeq a(t) \quad , \quad t_i \simeq \frac{2}{3H_i\sqrt{\Omega_i}} , \tag{32}$$

and

$$H(t) \simeq H_i\sqrt{\Omega_i}\frac{t_i}{t} . \tag{33}$$

With

$$\Omega_i \equiv 1 - \epsilon_i \simeq 1 - \frac{1-\Omega_0}{\Omega_0(1+z_i)} , \tag{34}$$

it follows

$$a^2(t)\dot{b}(t) \simeq \frac{3f}{2}H_i^2\Omega_i t \simeq \frac{3}{2}H_i^2 t \, g_i(t) ; \tag{35a}$$

$$g_i(t) \equiv \Omega^{0.6}(t)(1-\epsilon_i) \simeq \Omega^{0.6}(t)\left[1 - \frac{1-\Omega_0}{\Omega_0(1+z_i)}\right] . \tag{35b}$$

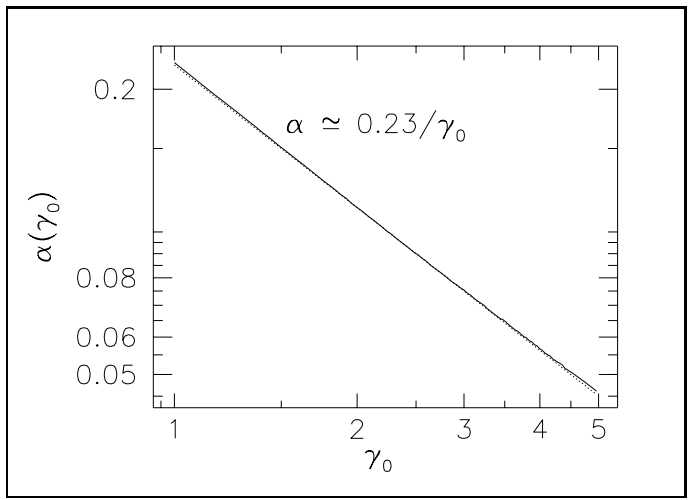

**Fig. 1.** The exponent $\alpha(\gamma_0)$ introduced in Eq.(38, solid line). The dotted line displays the approximation $\alpha \simeq (0.23/\gamma_0)$, which is seen to provide an excellent fit

If a perturbation collapses early, the density parameter $\Omega(t)$ at collapse time will be close to unity and depend only very weakly on $t$. Then, Eq.(35a) shows that in this case the spin parameter grows proportional to $t$ as long as the linear approximation provides a viable description.

We now insert for $t$ the time when a homogeneous sphere of density contrast $\delta_i$ achieves its maximum extension; this is given by

$$H_i t_{\max} = \frac{\pi}{2\left(\frac{5}{3}\delta_i - \epsilon_i\right)^{3/2}} \ ; \tag{36}$$

see, e.g., Bartelmann et al. [1993, Eq.(26)]. Inserting into Eq.(35a), we end up with

$$\frac{a^2(t_{\max})\dot{b}(t_{\max})}{H_i} \simeq \frac{3\pi}{4} g_i(t_{\max})\gamma_i^{-3/2} \ . \tag{37}$$

It will later on turn out favourable to 'rephrase' the $t_{\max}$-dependency of the function $g_i$. Of course, the most straightforward procedure would be to just insert the (analytically known) functions $a(t)$ and $\dot{b}(t)$ into Eq.(29) and avoid any approximations. However, since we are aiming on the formation of galaxies or rich clusters, we suspect the collapse time $t_{\max}$ to be early enough for the Einstein-de Sitter limit to be a reasonable approximation. Then, the further treatment can be considerably simplified.

We start with the ansatz

$$g_i(t_{\max}) = \Omega_0^\alpha \tag{38}$$

and determine the coefficient $\alpha$ by performing a least-square fit of the approximation (37) to the exact function $a^2(t)\dot{b}(t)$. The so-determined $\alpha$ will depend on the amplitude of the density perturbation. For large overdensities, we expect $\alpha \to 0$, since then the density parameter at the turnaround time will be very close to unity, while for low overdensities, $\alpha$ should approach 0.6.

Figure 1 displays $\alpha$ as a function of $\gamma_0 \equiv \gamma_i b(z)|_{z=0}$. The figure clearly shows that, in the range considered, $\alpha$ is very well approximated by a power law,

$$\alpha(\gamma_0) \simeq \frac{0.23}{\gamma_0} \ . \tag{39}$$

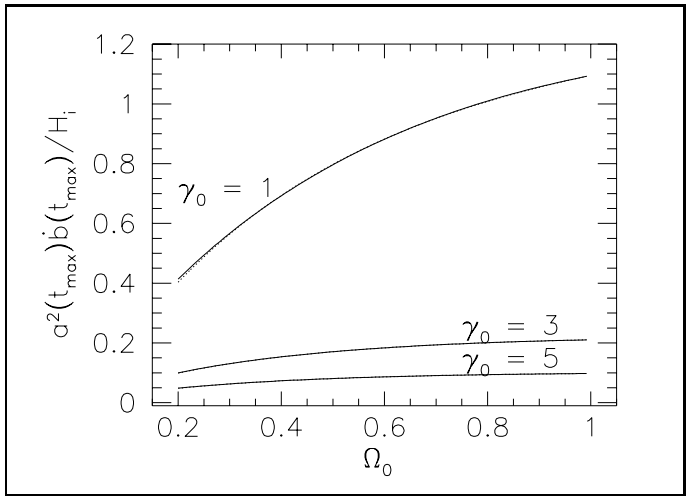

**Fig. 2.** The factor $a^2(t_{\max})\dot{b}(t_{\max})$ occurring in the expression (29) for the spin parameter, scaled by $H_i$. The solid line gives the exact (numerically determined) value, the dotted line the approximation of Eq.(37) for $\gamma_0 \in \{1, 3, 5\}$, which can be considered typical for fluctuations evolving into clusters of galaxies or galaxies (see text)

This confirms that for high $\gamma_0$ and thus large density perturbations, $\alpha \to 0$; e.g., for $\gamma_0 = 5$ we have $\alpha \simeq 0.05$. Thus, we can write with an error of about 0.1%

$$g_i(\Omega_0, \gamma_0) \simeq \Omega_0^{0.23/\gamma_0} \; . \tag{40}$$

Moreover, we compare in Fig.2 the exact function $a^2(t)\dot{b}(t)$ with the approximation (37), where $g_i$ is taken from (40); curves for $\gamma_0 \in \{1, 3, 5\}$ are displayed. The figure shows that the approximation (37) to the exact behaviour of the time-dependent term is excellent.

Thus, the $\Omega_0$ dependence is rather weak, $\alpha = 0.05\ldots 0.1$ being typical. That means, that for galaxy-sized perturbations, the spin parameter should practically be independent of $\Omega_0$. Even for much larger fluctuations (e.g., fluctuations on cluster scale), with $\gamma_0 \simeq 1$, the dependence $\propto \Omega_0^{0.23}$ is still weak. For consistency, we now consider $g_i$ as a function of $\epsilon_i$ and $\delta_i$, $g_i(\epsilon_i, \delta_i) = g_i[\Omega_0(\epsilon_i), \gamma_0(\delta_i, \epsilon_i)]$. Introducing this expression into Eq.(37), the square of the spin parameter becomes

$$\lambda^2 \simeq \frac{27\pi^2}{1000} \frac{g_i^2(\epsilon_i, \delta_i)}{\gamma_i^2} \frac{\mathrm{F}(\theta, k)}{\sin \theta} \frac{\sum_{j>k}(\mu_j - \mu_k)^2 \Phi_{,jk}^2}{\mu_1 \sqrt{\mu_2 \mu_3}} \; . \tag{41}$$

Note that $\lambda^2$ depends on a sum of terms $\propto \Phi_{,jk}^2/\gamma_i^2$. In case of an Einstein-de Sitter universe, this term converges to a finite value for $\delta_i \to 0$ because of Poisson's equation. If, however, $\epsilon_i > 0$, it diverges for $\delta_i \to \frac{3}{5}\epsilon_i$. This can be understood as follows: In an open universe, density perturbations with $\frac{5}{3}\delta_i < \epsilon_i$ do never collapse but 'freeze in' at $z \simeq 1/\Omega_0 - 1$. Nevertheless, the tidal field is present, i.e., the tidal torques are still acting, the angular momentum grows with time. Within linear theory, such a non-collapsing 'dark halo' would acquire an infinite amount of angular momentum. Therefore, we expect that for lately formed objects, the spin parameter should be higher. On the other hand, the factor $g_i = \Omega_0^{0.23/\gamma_0}$ suppresses the angular momentum due to the larger 'Hubble friction' in an open universe. This becomes more important with decreasing $\gamma_0$, because for high $\gamma_0$, most of the gravitational evolution is finished when the cosmic evolution is still indistinguishable from that of a flat universe. For lower $\gamma_0$, a noticible fraction of the dynamical time is spent during the phase when $\Omega$ significantly deviates from unity, where the 'Hubble friction' becomes more important. We will see later on that these two effects balance very well for objects with $\gamma_0 \gtrsim 1$ (see Fig.5).

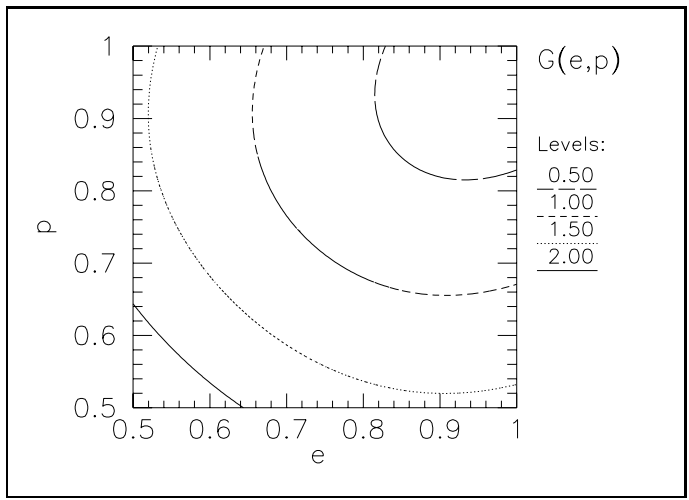

**Fig. 3.** Contour plot of the shape-dependent function $G(e,p)$ for $(e,p) \in [0.5, 1]$

## 3 Statistics

In the following section, the statistical properties of the spin parameter will be considered within the framework of Gaussian random fields. A strictly homogeneous Gaussian random field is entirely determined by its power spectrum $\langle \delta_k^2 \rangle = P(k)$. Its variance $\sigma_i$, given by

$$\sigma_i^2 \equiv \int \frac{d^3k}{(2\pi)^3} P(k, t_i) W_X(k) \;, \qquad (42)$$

is a measure for the mean peak height at the time $t_i$. $W_X$ is the window function smoothing out fluctuations on scales smaller than the typical extension $X$ of the considered volume. With $\sigma_0$ we will design the variance at $z = 0$, extrapolated from $z_i$ via Eq.(2).

### 3.1 Ensemble average

Averaging over ensembles, and assuming that the quantities determining the shape of the ellipsoid are statistically independent of $\Phi_{,jk}$ and $\delta_i$, we obtain

$$\langle \lambda^2 \rangle = \frac{27\pi^2}{1000} \left\langle \frac{g_i^2(\epsilon_i, \delta_i)\Phi_{,12}^2}{\gamma_i^2} \right\rangle \left\langle \frac{\mathrm{F}(\theta,k)}{\sin\theta} \frac{\sum_{j>k}(\mu_j - \mu_k)^2}{\mu_1\sqrt{\mu_2\mu_3}} \right\rangle \;; \qquad (43)$$

in this equation, it was further assumed that, on average, $\langle \Phi_{,jk} \rangle$ is the same for all $i \neq j$, so that $\Phi_{,jk}$ can be replaced by $\Phi_{,12}$. Hence, the spin parameter separates into a term dependent only on the (initial) geometry of the ellipsoid,

$$\langle G^2(e,p) \rangle \equiv \left\langle \frac{\mathrm{F}\left(\arccos(p), \sqrt{\frac{1-e^2}{1-p^2}}\right)}{ep\sqrt{1-p^2}} \left[3(1 + e^4 + p^4) - (1 + e^2 + p^2)^2\right] \right\rangle \;, \qquad (44)$$

and a term dependent on the cosmological density parameter (via $\epsilon_i$) and the statistical properties of the potential $\Phi$, which are incorporated in the perturbation spectrum,

$$\tilde{K}^2(\sigma_i, \epsilon_i) \equiv \left\langle \frac{g_i^2(\epsilon_i, \delta_i)\Phi_{,12}^2}{\gamma_i^2} \right\rangle \;. \qquad (45)$$

Figure 3 displays contours of the shape-dependent function $G(e,p)$.

The assumption that the shape of the volume $\Upsilon$ is statistically independent of the surrounding tidal field $\Phi_{,jk}$ can be motivated by the following argument. Let the tidal field $\Phi_{,jk}$ be decomposed into two parts $\Phi_{,jk}^{\rm int}$ and $\Phi_{,jk}^{\rm ext}$. Sources for $\Phi_{,jk}^{\rm int}$ and $\Phi_{,jk}^{\rm ext}$ are the matter distributions inside and outside $\Upsilon$, respectively. The principal-axis system of $\Phi_{,jk}^{\rm int}$ has to be aligned with $I_{jk}$ because otherwise the homoeoid would be spun up by its own tidal field. The alignment between the external and the internal tidal fields, if present, should be weak for the assumption of statistical indepence between $\Phi_{,jk}^{\rm ext}$ and $\Phi_{,jk}^{\rm int}$ to be valid. $N$-body simulations by Barnes & Efstathiou (1987) support this view, in that they exhibit only a weak tendency for the principal axes (or spin vectors) of neighbouring haloes to be aligned. In any case, the assumption of statistical independence tends to overestimate the growth of angular momentum, because if the principal-axis system of $\Upsilon$ tends to be partially aligned with the external tidal field, the angular momentum gain of $\Upsilon$ is reduced.

### 3.2 Eigenvalue distribution of $\Phi_{,jk}$

We now statistically evaluate the cosmological factor $\langle g_i^2 \Phi_{,12}^2 / \gamma_i^2 \rangle$. This evaluation is easiest performed rotating first into the principal-axis system of the tidal tensor $\Phi_{,jk}$ and applying a random rotation to the result afterwards. Averaging over the three Euler angles of this rotation, we will obtain the desired relation in Eq.(51) below. Since $\gamma_i$ depends only on the trace of $\Phi_{,jk}$, it is unaffected by rotations; therefore, the principal-axis system of $g_i^2 \Phi_{,jk}^2 / \gamma_i^2$ is identical with that of $\Phi_{,jk}$.

Since $\Phi_{,jk}$ is symmetric, there exists a local coordinate transformation $T(\alpha, \beta, \gamma)$ dependent on the three Euler angles $(\alpha, \gamma) \in [0, 2\pi)$, $\beta \in [0, \pi)$ such that

$$T^{-1}(\alpha, \beta, \gamma) \, (\Phi_{,jk}) \, T(\alpha, \beta, \gamma) = \mathrm{diag}(\varphi_1, \varphi_2, \varphi_3) \,. \tag{46}$$

For a Gaussian random field $\Phi$, the probability distribution of the eigenvalues $\varphi_j$ of $\Phi_{,jk}$ is known to be

$$\begin{aligned} p(\phi_1, \phi_2, \phi_3) d^3\phi = {} & \frac{15^3}{8\pi\sqrt{5}} \, |(\phi_3 - \phi_1)(\phi_3 - \phi_2)(\phi_2 - \phi_1)| \\ & \times \exp\left\{-\frac{3}{2} \left[2(\phi_1^2 + \phi_2^2 + \phi_3^2) - (\phi_1\phi_2 + \phi_1\phi_3 + \phi_2\phi_3)\right]\right\} d^3\phi \,; \end{aligned} \tag{47}$$

(Doroshkevich 1970, Bartelmann & Schneider 1992). We have introduced $\phi_j \equiv (\varphi_j/\sigma_i)$ with $\sigma_i$ defined by the perturbation spectrum of the density contrast $P(k)$ [Eq.(42)]. The initial density contrast is given by Poisson's equation,

$$\delta_i = \Delta\Phi = \mathrm{tr}\,(\Phi_{,jk}) = \sigma_i \sum_{i=1}^{3} \phi_j \,. \tag{48}$$

As motivated in section 3.1, the principal-axis system of the tidal tensor $\Phi_{,jk}$ can be assumed to be independent of the principal-axis system of the inertial tensor $I_{jk}$. Then, the Euler angles $(\alpha, \beta, \gamma)$ between the two systems assume random values in their intervals of definition. Consider an arbitrary diagonal matrix $M_{jk} = \mathrm{diag}(m_1, m_2, m_3)$ which is arbitrarily rotated by a rotation matrix $T(\alpha, \beta, \gamma)$,

$$M'_{jk}(m_1, m_2, m_3, \alpha, \beta, \gamma) = T(\alpha, \beta, \gamma) \, (M_{jk}) \, T^{-1}(\alpha, \beta, \gamma) \,. \tag{49}$$

It can then be shown by straightforward, but rather tedious algebra that, averaging over the Euler angles,

$$\left\langle (M'_{11})^2 \right\rangle = \frac{1}{15} \left( 3 \sum_{j=1}^{3} \langle m_j^2 \rangle + 2 \sum_{j \neq k} \langle m_j m_k \rangle \right) ,$$
$$\left\langle (M'_{12})^2 \right\rangle = \frac{1}{15} \left( \sum_{j=1}^{3} \langle m_j^2 \rangle - \sum_{j \neq k} \langle m_j m_k \rangle \right) . \tag{50}$$

If the $m_j$ have the same statistical properties for all $j$, and if $\langle m_j m_k \rangle = 0$ for $j \neq k$, as is the case for the $\varphi_j$, Eq.(50) can be simplified to

$$\left\langle (M'_{jj})^2 \right\rangle = \frac{3}{5} \langle m_1^2 \rangle ,$$
$$\left\langle (M'_{jk})^2 \right\rangle = \frac{1}{5} \langle m_1^2 \rangle ; \tag{51}$$

the same result could have been obtained also by considering invariants of $M_{jk}$. For example, we obtain

$$\langle \phi_1^2 \rangle = \frac{1}{3} \Rightarrow \langle \Phi_{,11}^2 \rangle = \frac{1}{5} , \quad \langle \Phi_{,12}^2 \rangle = \frac{1}{15} ; \tag{52}$$

these results agree with those given in Bardeen et al. [1986, Eq.(A1)].

### 3.3 Conditional probability

With Eq.(51), we can write

$$\left\langle \frac{g_i^2(\epsilon_i, \delta_i) \Phi_{,12}^2}{\gamma_i^2} \right\rangle = \frac{9}{125} \int d^3\phi \, p(\phi_1, \phi_2, \phi_3) \left[ \frac{g_i^2(\epsilon_i, \delta_i) \phi_1^2}{\left( \sum \phi_j - \omega_i \right)^2} \right] , \tag{53}$$

where $\omega_i \equiv (3/5)(\epsilon_i/\sigma_i)$. Within this notation, the effective density contrast is given by $\gamma_i = \frac{5}{3}\sigma_i (\sum \phi_j - \omega_i)$.

The integral in Eq.(53) has to be evaluated for sufficiently large values of $\delta_i$, i.e., restricting to such density perturbations which collapse at moderate to high redshifts. We therefore choose

$$\delta_i - \frac{3}{5}\epsilon_i \geq \nu \sigma_i \Rightarrow \sum_{i=1}^{3} \phi_j - \omega_i \geq \nu , \tag{54}$$

where $\nu \gtrsim 1$ appears as a reasonable choice. Note that because of Poisson's equation, $\Delta \Phi = -\delta$, the restriction (54) on the density contrast also restricts the tidal field. The integral in Eq.(53) then has to be replaced by

$$\frac{\int d^3\phi \, p(\phi_1, \phi_2, \phi_3) \left[ \frac{g_i^2(\epsilon_i, \delta_i) \phi_1^2}{\left( \sum \phi_j - \omega_i \right)^2} \right] \mathrm{H}\left( \sum \phi_j - \omega_i - \nu \right)}{\int d^3\phi \, p(\phi_1, \phi_2, \phi_3) \mathrm{H}\left( \sum \phi_j - \omega_i - \nu \right)} \equiv K^2(\sigma_i, \epsilon_i, \nu) , \tag{55}$$

where $\mathrm{H}(x)$ is the Heaviside step function; Eq.(55) expresses Bayes' formula for conditional probabilities. Combining Eqs.(43), (53), and (55), we finally obtain

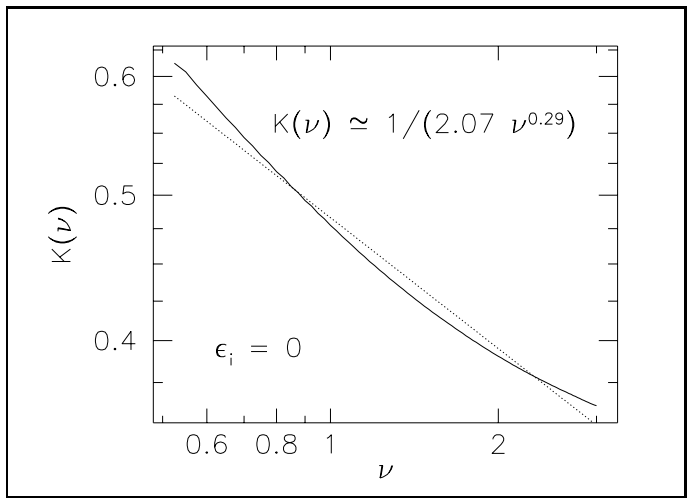

**Fig. 4.** The function $K(\sigma_i, \epsilon_i, \nu)$ for the Einstein-de Sitter case, $\epsilon_i = 0$. In this case, $K$ does only depend on $\nu$. The solid line gives $K(\nu)$, the dotted line shows the power-law fit of Eq.(58)

$$\langle \lambda^2 \rangle = \lambda_0^2 K^2 (\sigma_i, \epsilon_i, \nu) G^2(e, p) , \qquad (56)$$

with

$$\lambda_0^2 = \frac{3^5 \pi^2}{50^3} \simeq 0.02 . \qquad (57)$$

As an example, Fig.4 shows the function $K(\sigma_i, \epsilon_i, \nu)$ for the Einstein-de Sitter case, i.e., for $\epsilon_i = 0$. Then, $K$ does no longer depend on $\sigma_i$, and is merely a function of $\nu$. As seen in the figure, $K(\nu)$ is reasonably fit by a power law in the range of $\nu$ given; we find with an error of $\lesssim 5\%$

$$K(\nu) \simeq \frac{1}{2.07 \, \nu^{0.29}} \quad \text{for} \quad \nu \in [0.5 \ldots 3] . \qquad (58)$$

For $\nu = 1$, we read off $K \simeq 0.48$. Since $G(e, p) \simeq 1$ for reasonable values of $e$ and $p$, we have

$$\sqrt{\langle \lambda^2 \rangle} \simeq 0.07 , \qquad (59)$$

which is in excellent agreement with numerically determined spin parameters. The small value of $\lambda$ supports the neglection of the rotational energy in Sect.2.3.

### 3.4 Dependence on the density parameter

We have defined $\epsilon_i = 1 - \Omega_i$, hence,

$$\epsilon_i = \frac{1 - \Omega_0}{\Omega_0(1 + z_i)} ; \qquad (60)$$

cf. Eq.(34). Since $\sigma_i$ and $\sigma_0$ are related by

$$\sigma_0 = \frac{1 + \tilde{\Omega} z_i}{1 + \tilde{\Omega} z} \sigma_i , \qquad (61)$$

with $\tilde{\Omega}$ from Eq.(4), the quantity $\omega_i$ is given by

$$\omega_i \equiv \frac{3}{5} \frac{\epsilon_i}{\sigma_i} \simeq \frac{3}{\sigma_0} \frac{1 - \Omega_0}{2 + 3\Omega_0} , \qquad (62)$$

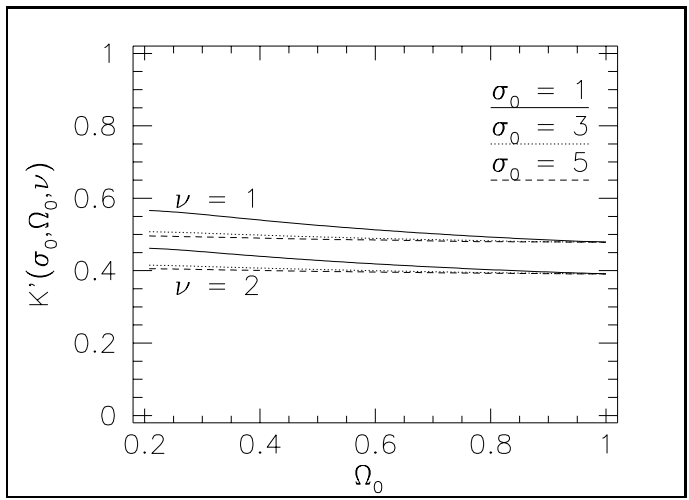

**Fig. 5.** The function $K'(\sigma_0, \Omega_0, \nu)$ for $\nu = 1, 2$ as a function of $\Omega_0$. The solid (dotted, dashed) line is for $\sigma_0 = 1$ (3,5, respectively)

where we have used $z_i \gg 1$. In terms of $\sigma_0$ and $\Omega_0$, we write

$$K'(\sigma_0, \Omega_0, \nu) \equiv K[\sigma_i(\sigma_0, \Omega_0), \epsilon_i(\Omega_0), \nu] . \tag{63}$$

In Fig.5, we show the dependence of $K'$ on $\sigma_0$ and $\Omega_0$ for $\nu = 1$. As expected, the dependency is very weak. Note, however, that for small $\Omega_0$ and $\sigma_0$, $K'$ is systematically higher. This reflects the fact that objects collapsing late in a low-density universe are exposed to the tidal field for a longer period and can, therefore, acquire higher angular momentum.

### 3.5 Probability distribution of the spin parameter

The probability distribution for the spin parameter can also be derived in a straightforward way. Consider the function $K'(\sigma_0, \Omega_0, \nu)$ introduced in Eqs.(55) and (63) as an average over the relevant ranges of the $\phi_j$. Since, as demonstrated in Figs.4 and 5, the dependence of $K'$ on both $\nu$ and $\Omega_0$ is weak, we restrict ourselves to the Einstein-de Sitter case ($\Omega_0 = 1$) and to $\nu = 1$.

Then, the statistical properties of $\lambda$ are uniquely determined by the probability distribution $p(\phi_j)$ for the eigenvalues of $\Phi_{,jk}$, and thus the probability distribution $\bar{p}(\lambda)$ for the spin parameter can be derived using Eq.(47) and transforming variables; it is displayed in Fig.6 together with the cumulative distribution

$$P(\lambda) \equiv \int_\lambda^\infty d\lambda' \bar{p}(\lambda') . \tag{64}$$

In the right panel of Fig.6, we indicate the region where

$$10\% \leq P(\lambda) \leq 90\% ; \tag{65}$$

in terms of $\lambda$, this interval is

$$0.02 \leq \lambda \leq 0.11 . \tag{66}$$

This is in surprisingly good agreement with results from numerical simulations. Furthermore, the shape of the distribution function (Fig.6, left panel) is quite similar to that obtained by the numerical simulations of Warren et al. (1992, Fig.18): The probability distribution steeply rises for $\lambda \lesssim 0.025$, where the probability distribution reaches its maximum. Then the probability begins to decrease, relatively slowly for $\lambda \lesssim 0.1$, and rather steeply beyond. However, the maximum and the turnaround points in the numerical simulations tend to be at lower $\lambda$. Therefore, our approach tends to overestimate the spin parameter. Possible explanations are the neglect of the correlations between shape and tidal field or systematical effects resulting from our choice of $\nu$ and the ellipticities $(e,p)$, and the approximation of the turnaround time by the time of maximum expansion of a corresponding sphere.

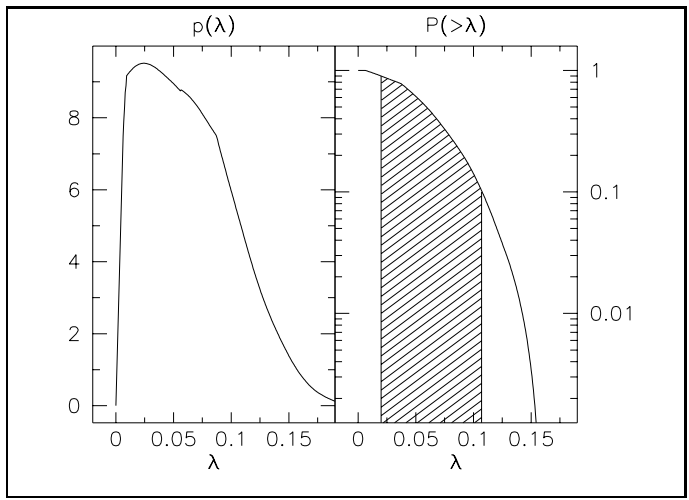

**Fig. 6.** Probability distribution (left) and cumulative probability distribution (right) for the spin parameter $\lambda$ in a Einstein-de Sitter universe with a peak height $\nu > 1$. The shadowed region marks the 80% interval, i.e., only 10% have lower and higher spin parameters, respectively.

## 4 Summary

We have investigated the spin parameter of dark haloes forming by gravitational collapse within the framework of first-order growth of angular momentum. Several approximations enter this derivation, these are:

1. The volume $\Upsilon$ is approximated by a homoeoid which is supposed to have the same inertial tensor as $\Upsilon$.
2. The energy of the volume (potential and kinetic) is approximated by taking the potential energy of a homoeoid at maximum expansion with the axis ratio of the homoeoid at initial time, while this axis ratio is generally depending on time.
3. The extension at maximum expansion of the homoeoid is approximated by the maximum radius of a sphere with the density contrast of the homoeoid.
4. The time-dependence of the spin parameter is evaluated in the high-redshift limit of the dynamics of the background cosmological model.
5. The principal-axis systems of $\Upsilon$ and the tidal field are taken to be statistically independent.

Assumption (5) is not obvious but seems to be well supported by numerical results. This assumption causes a slight overestimate of the spin parameter in the formalism above. This might be the reason for the slightly smaller spin parameters found in numerical simulations. Approximation (4) was shown in Sect.2 to be very well satisfied. Approximations (2) and (3) only affect the energy of the homoeoid, and the spin parameter depends only weakly on the energy. Additionally, Bartelmann et al. (1993) showed that the collapse history of a homoeoid is well approximated by the collapse timescale of a sphere, at least in a statistical sense. Approximation (1) is not crucial because the gain of angular momentum is controlled only by the inertial tensor of the mass distribution considered.

On the basis of these approximations, the spin parameter can be written

$$\sqrt{\langle \lambda^2 \rangle} \simeq \frac{\lambda_0}{2.07 \, \nu^{0.29}} \, G(e,p) \simeq 0.07 \, \nu^{-0.29} \, G(e,p) \,,$$
$$\lambda \in [0.02 \ldots 0.11] \, \nu^{-0.29} \, G(e,p) \quad \text{for} \quad \nu \in [0.5, 3] \,. \quad (67)$$

Its median value is $\bar{\lambda} \simeq 0.06$, and the limits on $\lambda$ are its 10 and 90 percentile values. In Eq.(67), $G(e,p) \simeq 1$ is determined by the shape of the dark halo, and $\nu$ is its initial density contrast $\delta_i$ in units of $\sigma_i$. The agreement with numerically determined spin parameters is astonishing, in particular, also the probability distribution of $\lambda$ (Fig.6, left panel) is well reproduced. The spin parameter increases linearly with time and approaches the value given by Eq.(67) after turnaround time, when the volume under consideration shrinks and makes it increasingly difficult for the external tidal field to spin it up.

As has been discussed above, the dependence of $\lambda$ on the cosmological density parameter is very weak (see Fig.5). Although the velocity factor $f \simeq \Omega^{0.6}(t)$ entered the derivation of $\lambda$, most of the 'interesting' density perturbations collapse sufficiently early for the Einstein-de Sitter limit to be a viable approximation; then, $\Omega$ is close to unity at turnaround time. When $\Omega_0$ is low, $\Omega_0 \gtrsim 0.2$, the normalization of the density perturbation spectrum today implies that perturbations which are now collapsed must have collapsed at higher redshift than in a high-density universe, because otherwise they could not have formed (see, e.g., Richstone et al. 1992, Bartelmann et al. 1993). Thus, at the formation epoch, $\Omega$ is still close to unity even if it is low today.

Additionally, Fig.5 shows that the dependence of $\lambda$ on the variance $\sigma_0$ of the linear density perturbation spectrum $P(k)$ is weak. This is due to the fact that both the tidal field and the density contrast of the volume considered are linear in $\sigma_0$, and since the ratio of these quantities enters the spin parameter, it does hardly depend on the spectral normalization. In turn, this implies that the spin parameter of collapsing objects does only very weakly depend on the type of perturbation spectrum inserted, since all perturbation spectra are normalized by observations on the spatial scales considered here. Thus, $\lambda$ is expected to be weakly, if at all, dependent on the cosmogonic scenario; this is a direct consequence of our smoothing the perturbation spectrum over the mass scale of interest. However, additional angular momentum can be gained after turnaround time by accretion of lower-mass objects, whose number density, and therefore the accretion rate, does depend on the power spectrum.

There is a mild dependence of $\lambda$ on the peak height of the perturbations considered; this peak height is measured in units of the spectral variance $\sigma_0$ by the parameter $\nu$. For higher $\nu$, the spin parameter decreases $\propto \nu^{-0.29}$. This behaviour reflects the fact that higher density perturbations reach their turnaround time earlier, so that their spin-up time is shorter. Since $\lambda$ is increasing linearly with time [cf. Eq.(35a)], a shorter spin-up time means a smaller spin parameter. This effect is mostly balanced because the tidal field in the vicinity of high density peaks is on average also enhanced due to the autocorrelation of high density peaks. Formally, this enters our derivation with condition (54), since, from Poisson's equation, the tidal field is restricted to high values together with the density contrast. Therefore, only combinations of high density fields with high tidal fields enter the ensemble average in Eq.(55).

While the numerical values for the spin parameter were previously known from numerical experiments, the virtue of the present derivation is that it explains the distribution of $\lambda$ in terms of the statistics of Gaussian random fields. Additionally, it explains why the spin parameter is insensitive to both $\Omega_0$ and the underlying cosmogonic model, be it CDM or HDM or some cocktail derived from these. We expect that the statistics of the spin parameter would considerably deviate from the results given here if the initial density field, and thus the tidal field, would significantly deviate from Gaussian statistics. Additionally, the derivation above shows that for the spin parameters of galaxies or clusters of galaxies, the angular momentum gain during the linear collapse phase of these objects is sufficient. In a universe with non-vanishing cosmological constant, the only difference would be that the expansion time-scale of the universe is enlarged, which means that the spin-up period would be longer; then, we would expect the spin parameter to increase.

*Acknowledgements.* We wish to thank Martin Hähnelt, Peter Schneider and Ewald Müller for their suggestions and for helpful discussions.